# QoS Based Dynamic Web Services Composition & Execution

Farhan Hassan Khan, M.Younus Javed, Saba Bashir
National University of Science & Technology
Rawalpindi, Pakistan

Aihab Khan, Malik Sikandar Hayat Khiyal
Fatima Jinnah Women University
Rawalpindi, Pakistan

*Abstract*---**The use of web services has dominated software industry. Existing technologies of web services are extended to give value added customized services to customers through composition. Automated web service composition is a very challenging task. This paper proposed the solution of existing problems and proposed a technique by combination of interface based and functionality based rules. The proposed framework also solves the issues related to unavailability of updated information and inaccessibility of web services from repository/databases due to any fault/failure. It provides updated information problem by adding aging factor in repository/WSDB (Web Services Database) and inaccessibility is solved by replication of WSDB. We discussed data distribution techniques and proposed our framework by using one of these strategies by considering quality of service issues. Finally, our algorithm eliminates the dynamic service composition and execution issues, supports web service composition considering QoS (Quality of Service), efficient data retrieval and updation, fast service distribution and fault tolerance.**

*Keywords*---*composition of services; dynamic composition; UDDI registry; web services.*

## I. INTRODUCTION

Web services are software applications that are available on the web and used for machine to machine interaction by using URI (Uniform Resource Identifier) on the distributed environment of internet. SOAP (Simple Object Access Protocol) messages are used for communication mechanism by using HTTP (Hyper Text Transfer Protocol) protocol. Each web service has an API (Application Program Interface) that can be accessed over the network and executes the web service at host location [1]. Every service provides a role, such as service provider, a requester or a broker. In other words, web services make possible the effective processing of machine readable information.

For business to business and enterprise level application integration, composition of web services plays an important role. Sometimes a single web service does not fulfill the user's desired requirements and different web services are combined through composition method in order to achieve a specific goal [2]. Service compositions reduce the development time and create new applications. Web services can be categorized in two ways on the basis of their functionality.1) Semantic annotation describes the functionality of web service and 2) functional annotation describes how it performs its functionality. WSDL (Web Services Description Language) is used for specification of messages that are used for communication between service providers and requesters [3].

There are two methods for web services composition [4,5,6]. One is static web service composition and other is automated/dynamic web service composition. In static web service composition, composition is performed manually, that is each web service is executed one by one in order to achieve the desired goal/requirement. It is a time consuming task which requires a lot of effort. In automated web service composition, agents are used to select a web service that may be composed of multiple web services but from user's viewpoint, it is considered as a single service [7].

The main interest of web service compositions is to give value-added services to existing web services and introduce automated web services. Also they provide flexibility and agility. There are few problems in dynamic web service composition as discussed in [8].

- First, the number of web services is increasing with time and it is difficult to search the whole repository for desired service in order to use it for the fulfillment of specific goal.
- Second, web services are dynamically created and updated so the decision should be taken at execution time and based on recent information.
- Third, different web service providers use different conceptual models and there is a need of one structure so that web services easily access each other without any technical effort.
- Forth, only authorized persons can access few of these web services.

The two approaches in web services composition are centralized dataflow and decentralized dataflow. In case of dynamic web services composition, both have advantages and some limitations. The limitation of centralized dataflow is that all component services must pass through a composite service. This results the bottleneck problem which causes the increase in throughput and response time. The disadvantage of decentralized dataflow is that as each web service directly shares data with web servers resulting in increasing the load at each node and delay in response time and throughput. The decentralized dataflow is very efficient in case of dynamic web services composition as it reduce the tight coupling between clients and servers by adding a middleware (UDDI (Universal Discovery, Description and Integration), WS (Web Service) coordination or WS transaction etc) [9]. In the proposed model we have used decentralized dataflow model which results in high throughput, minimum response time and latency.

Mostly automated composition techniques are interface based and functionality based. In interface based composition, on the bases of inputs and outputs through interfaces users get composite services and after composition desired results are achieved. The drawback of this approach is that functionality is not guaranteed, whereas in functionality based composition, with inputs and outputs user provides the formula that explains logic into interface information.

*A. Contribution*

The major contribution of this paper is that it presents a method for automated and dynamic web service composition by combination of interface based and functionality based approaches. It focuses on the data distribution issues, QoS issues and defines how execution problems can be avoided. This research also resolves the problems of decentralized dataflow and provides a framework that has minimum latency, maximum throughput and response time. This paper proposed a solution for researchers who are facing the problems of web service composition due to constant changes in input/output parameters, networking issues and independent nature of different web services.

Section 2 presents the introduction of existing dynamic web services composition techniques and highlights the advantages and disadvantages of these techniques. Section 3 presents a detailed overview of web services composition and dynamic web services composition. Section 4 describes the proposed framework and its working. Section 5 is concerned with proposed technique including methodology and algorithms. Implementation and evaluation is described in section 6. Finally, conclusion and future work is given in section 7.

## II. RELATED WORK

Incheon Paik, Daisuke Maruyama [2] proposes a framework for automated web services composition through AI (Artificial Intelligence) planning technique by combining logical combination (HTN) and physical composition (CSP (Constraint Satisfaction Problem)). This paper discusses the real life problems on the web that is related to planning and scheduling. It provides task ordering to reach at desired goal. OWL-S (Ontology Web Language-Semantic) and BPEL4WS (Business Process Execution Language for Web Services) are used for composition which removes the limitations of HTN (Hierarchical Task Network) that are lack of interactive environment for web services, lack of autonomy etc. Then the proposed model is compared with HTN according to web services invocation. It tackles the following given problems faced by planner alone; First, it does not deal with various web users requests for information. Second, it is inefficient for automated finding of solutions in given state space. Third, its maintenance is weak due to frequent user requests. The proposed framework provides intelligent web services for web users. It uses CSP which provides problem space for planning, scheduling and automation of desired task.

Faisal Mustafa, T. L. McCluskey [9] outlined the main challenges faced by automated web services composition that are related to distributed, dynamic and uncertain nature of web. The proposed model is semi-automatic and fixes some issues of dynamic web services composition that are listed as follows. First, repository has large number of web services and it is not possible to analyze and integrate them from repository. Second, updated web service information is required from repository when it is selected to fulfill the specific task. Third, multiple services are written in different languages and there is a need of conceptual model to describe them in a single service. The proposed technique has few drawbacks. First, if a server goes down then input/output issues may arise. Second, new uploaded information is not available in repository as it does not update its contents.

Pat. P. W. Chan and Michael R. Lyu [10] proposed the dynamic web service composition technique by using N-version programming technique which improves the reliability of system for scheduling among web services. If one server fails, other web servers provide the required services. Web services are described by WSDL (Web Services Description Language) and their interaction with other web services is described by WSCI (Web Services Choreography Interface). The composed web services are deadlock free and reduce average composition time. Also the proposed system is dynamic, as it works with updated versions without rewriting the specifications. At the end experimental evaluation and results are presented to verify the correctness of algorithm.

LIU AnFeng et al. [11] presents the technique based on web services interfaces and peer to peer ontology. It provides an overlay network with peer to peer technologies and provides a model for web services composition. The web services composition is based on domain ontology and Distributed Hash Table (DHT) is used for discovery and composition. The analysis shows that it is easy to understand because of loosely coupled due to the separation of interfaces from underlying details. The proposed model is based on ontology and service composition interface. It provides QoS based composition, fast composition rate, fault tolerant and efficient discovery.

Kazuto Nakamura, Mikio Aoyama [12] proposed a technique for dynamic web service composition that is value based and provides composed web services based on QoS. Value meta-model and its representation language VSDL (Value-based Service Description Language) are presented. Values are used to define quality of web services. Value added service broker architecture is proposed to dynamically compose the web services and value-meta model to define relationship among values. The results explained that resultant composite services can provide more values of quality of contents as compared to previous discovered web services.

Although a number of dynamic web services composition techniques have been introduced, there is a need of dynamic approach to handle the large number of increasing web services and their updation in repositories. In this paper, we aim at providing an automated, fault tolerant and dynamic web services composition framework.

### III. PRELIMINARIES

This section gives some basic information about web services composition, automated web services composition and actors involved in dynamic web services composition.

#### A. Web Services Composition

Web services are distributed applications. The main advantage over other techniques is that web services can be dynamically discovered and invoked on demand, unlike other applications in which static binding is required before execution and discovery. Semantic and ontological concepts have been introduced to dynamically compose web services in which clients invoke web services by dynamically composing it without any prior knowledge of web services. Semantic and ontological techniques are used to dynamically discover and compose at the same time (run time).

#### B. Automated Web services Composition

The automated web service composition methods generate the request/response automatically. Most of these methods are based on AI planning. First request goes to Translator which performs translation from external form to a form used by system, and then the services are selected from repositories that meet user criteria. Now the Process Generator composes these services. If there are more than one composite service that meet user criteria then Evaluator evaluates them and returns the best selected service to Execution Engine. The results are returned to clients (Requester). There should be well defined methods and interfaces through which clients interact with the system and get the response. The generalized dynamic composition framework is shown in Fig 1. [6]

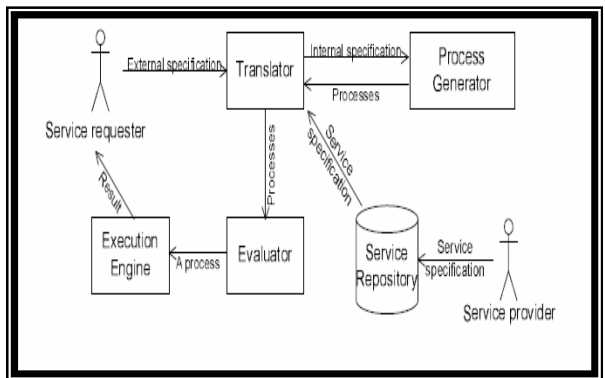

Fig 1: Automated Web Services Composition

### IV. PROPOSED FRAMEWORK

The proposed framework consists of following components as shown in Fig 2.

1. *Service Registration:* It is the process of specification of web services to the system. New services are registered in registry through service registration process. There are several registries that different companies (service providers) maintain and all of them are synchronized after regular interval.
2. *Service Request:* Clients that need a particular service send request through service request module.
3. *Translator:* The purpose of translator is to translate request/response from one form to another. We use translator so that all of the services are registered from external form to a form used by system and vice versa.
4. *Web Server:* Registries are hosted on web server on World Wide Web. Services exchange data directly with various databases and web servers, that implements decentralized data flow.
5. *Evaluator:* The evaluator evaluates selected web services on the basis of interface based and functionality based rules and returns the best selected service based on specified criteria.
6. *Web:* In proposed framework, web is World Wide Web network where all service providers register their web services in UDDI registries. If desired web services are not found in repository or database then matching engine will search them from UDDI registries and save it in database for current and future use.
7. *Composer:* The composer composes the selected component services in order to make a single desired web service.
8. *Matching Engine:* The purpose of matching engine is to match the user's request from the web services database. If match is found, it returns results back to web server. If not then select the web services from web, store/update them in database and then return results back to requested composer.
9. *Service Registry:* The service registries are used to register the web services by web service providers. Also they are used to request the user's desired web services. Each registry has the references of web services that are actually hosted on service repositories.

### V. PROPOSED TECHNIQUE

#### A. Methodology

The methodology of proposed model is given as:

1. The web services are registered in registries.
2. Service requester send request for service.
3. Translator converts the query into a form used by internal system.

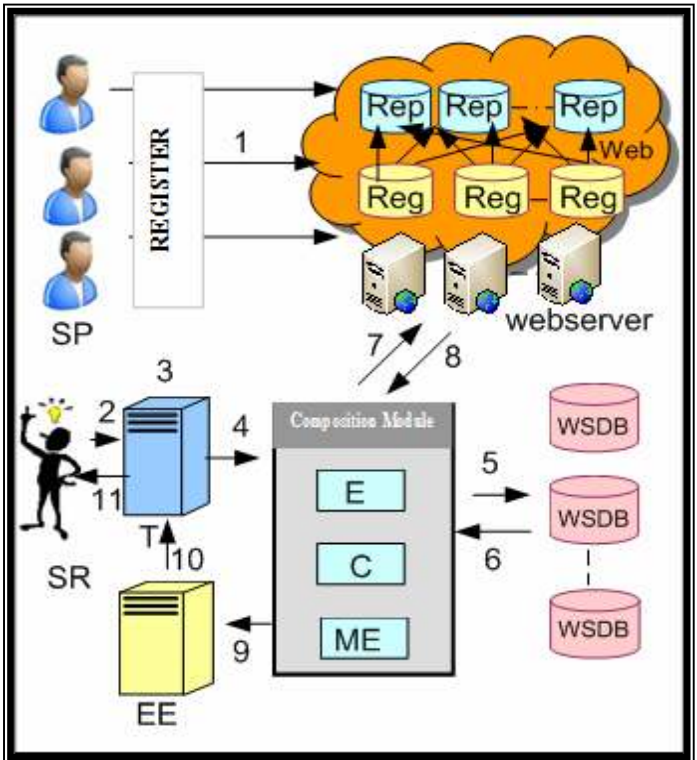

Fig 2: Proposed framework of web services composition

The abbreviations used in framework diagram are:
SP= Service Provider     C= Composer
SR= Service Requestor     Reg= UDDI Registry
T= Translator     Rep= Repository
EE= Execution Engine     E= Evaluator
ME= Matching Engine

4. The request arrives at composition module. Matching Engine checks for the requested service from WSDBs. If it finds the desired interface base service composition then it sends results to Evaluator.
5. Evaluator evaluates these selected web services in two steps. In first step it evaluates the web services on the basis of interface based search, whereas in second step it performs the evaluation on basis of functionality based rule. After evaluation it sends selected services to composer. The purpose of composer is to compose these component web services. Multiple WSDBs are introduced, so that if one goes down then we can make use of other databases. A timestamp (aging) is maintained with each URI in WSDB. If the request arrives before the expiration of that time then it looks for the service in WSDB.
6. If Matching Engine does not find requested service composition from web services database then it start searching from web.
7. The web services are searched from multiple registries and results are returned to Evaluator. Matching Engine also saves their references in WSDB with aging factor. The purpose of aging is that it maintains the updated information about web services as the contents are refreshed each time when aging time expires.
8. Evaluator evaluates these web services based on interface based and functionality based rules.
9. Composer composes these evaluated resultant services and sends result to Execution Engine. Execution Engine executes these web services and through translator results are sent back to requester.

*B. Pseudo Code*

The pseudo code of proposed framework is given as:

```
Algorithm: Web services composition
Input: Request for web service
Output: Composed service

Web services registered in registries;
Translator translates input query;
User enters request for web service;
Request goes to Matching engine;
    Matching engine search services with valid
    timestamp from WSDB;
    Matching engine select valid services;
    Evaluator evaluates above selected
    services based on interface based and
    functionality based rules;
    Composer composes evaluated services;
    Composer sends result to matching
    engine;
    Matching engine sends result Execution
    Engine;
    Execution engine execute these services
    and results are sent to requestor through
    translator;
 If no service found or timestamp expired
    Matching engine search service from
    Web UDDI registries;
    Select matched services;
    Evaluator evaluates above selected
    services based on interface based and
    functionality based rules;
    Composer composes evaluated services;
    Store matched services in WSDB with
    timestamp;
    Composer sends result to matching
    engine;
    Matching engine sends result Execution
    Engine;
    Execution engine execute these services
    and results are sent to requestor through
    translator;
```

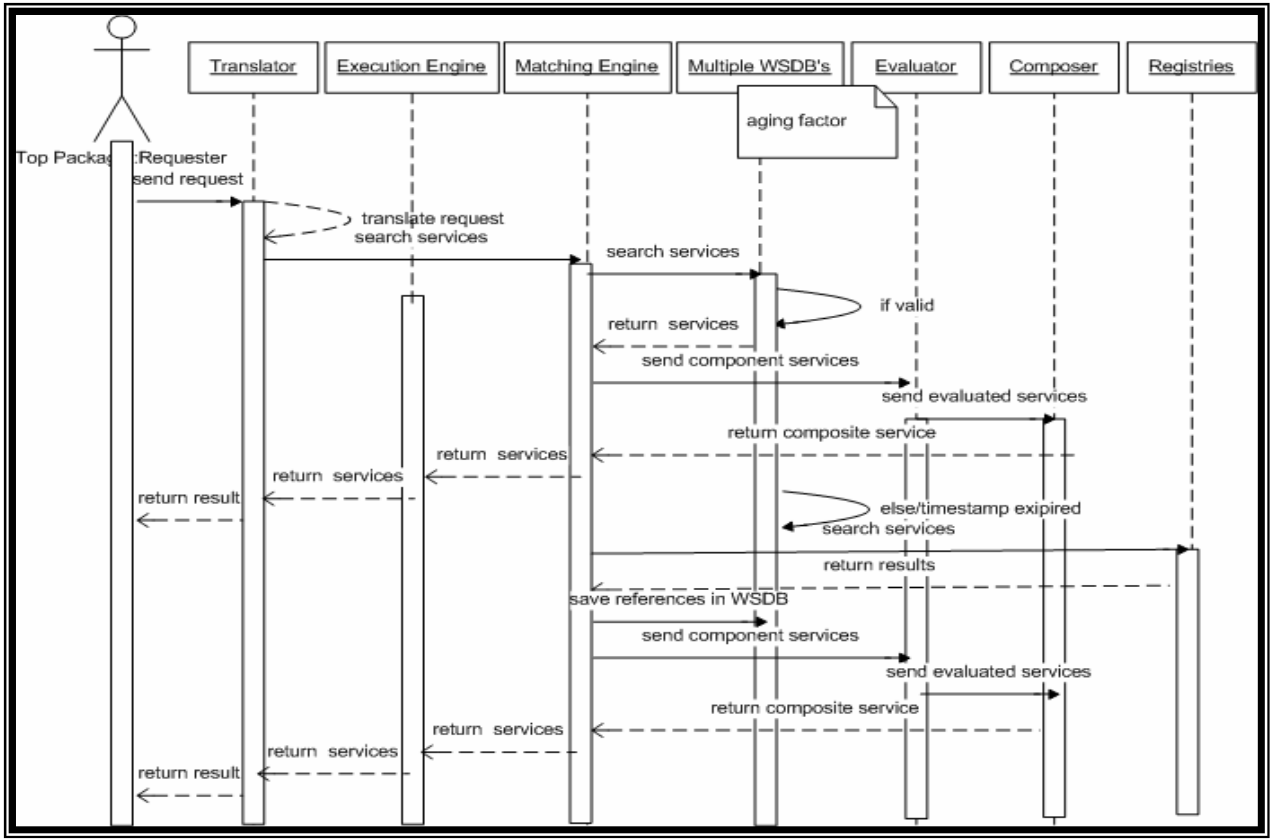

Fig 3: Sequence diagram of proposed framework

## VI. IMPLEMENTATION & EVALUATION

JAVA programming language is used to implement the proposed technique. We have used Apache JUDDI (Java implementation of Universal Description Discovery and Integration version 3), which is java implementation of UDDI registries. RUDDI is used to access the JUDDI. Service providers can perform various operations in UDDI registries like save, edit and delete services and businesses by using RUDDI. JAXR (Java API for XML Registries) and UDDI4J also provides the java API to access UDDI registries but the drawback is that JAXR is Java API to access various types of XML registries like UDDI, ebXML and others. It is not specific to UDDI registry. Whereas UDDI4J supports only UDDI version 2. We have also used WSDL4J (Web Services Description Language for Java Toolkit) which is Java API used to access WSDL files. WSIF (web services invocation framework) is used for invocation of multiple services from different repositories. Service providers register their web services in UDDI registries. Requesters requests for web service and translator translates it. Then matching engine search the requested service from web services database. The valid services are returned to evaluator and send the evaluated services to composer. Composer sends the composed service to execution engine and returns the resultant service to requester through translator. If web services are not found in WSDB then matching engine will search from UDDI registries. The matched services are returned to evaluator which evaluates them and after composition resultant service is passed to execution engine. Execution engine execute these services and results are sent to requestor through translator.

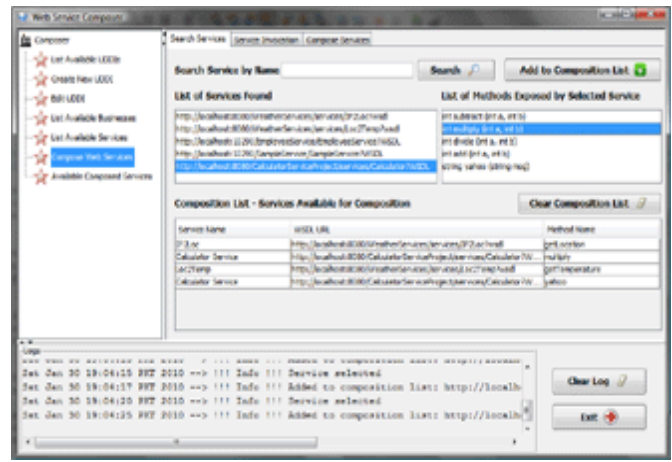

Fig 4: GUI of proposed web services composition application

*A. Performance Evaluation and Analysis*

In the proposed technique, UDDI registries provide efficient access. The efficiency is further improved since composite services once discovered, are stored in local WSDBs and till the timer expires these composite service can be used locally. Replicated WSDBs make the process reliable. Associated timestamp in WSDBs makes it possible to use any new services that become available on the registry.

The purpose of our algorithm is to invoke and compose web services efficiently. The effectiveness of algorithm is evaluated by calculating the method exposed time and service composition time. We have taken different number of methods exposed and calculated their respective time. We have also taken different number of composed services and calculated their respective execution time. It is clearly shown in Figure 5. Figure 6 shows the different number of methods exposition time.

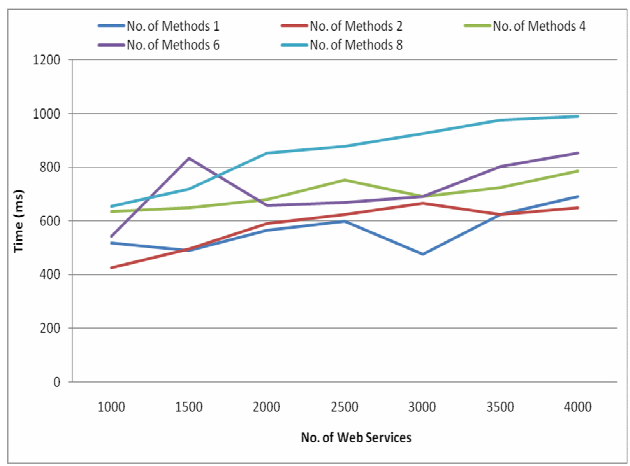

Fig 5: Efficiency of proposed algorithm

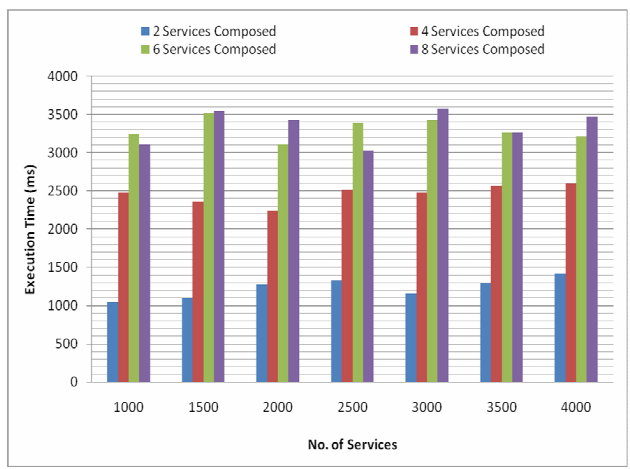

Fig 6: Efficiency of proposed algorithm

## VII. CONCLUSION & FUTURE WORK

The research lies in the field of dynamic web services composition selection. In this paper we discussed the main problems faced by dynamic web services composition. This paper proposed the dynamic web services composition algorithm to solve the composition issues related to data distribution, reliability, availability and QoS. It presented a framework in which multiple repositories and WSDBs have been introduced in order to make system more reliable and ensure data availability. By using multiple registries, data availability is guaranteed whereas by using aging factor users can retrieve up to date information. The proposed system is fault tolerant, reliable, performs fast data retrieval and Quality of service based. In future, the framework can be extended by crawling the web for searching web services instead of querying the UDDI registries. We will also be looking into deeper details of every component of the framework to ensure better and efficient composition.